\documentclass[journal]{IEEEtran}

\usepackage{graphicx}
\usepackage{color}
\usepackage{placeins}
\usepackage{float}
\usepackage{tabularx,colortbl}
\usepackage{amssymb}
\usepackage{amsthm}
\usepackage{cite}
\usepackage{amsmath}

\usepackage{caption2}

\usepackage{algorithm}
\usepackage{algpseudocode}

\theoremstyle{plain}
\newtheorem{thm}{Theorem}

\theoremstyle{plain}
\newtheorem{rem}{Remark}

\IEEEoverridecommandlockouts

\begin{document}

\title{Rate-Splitting for CF Massive MIMO Systems With Channel Aging}

\author{Jiakang~Zheng, Jiayi~Zhang,~\IEEEmembership{Senior Member,~IEEE}, Hongyang~Du, Dusit~Niyato,~\IEEEmembership{Fellow,~IEEE}, Dong~In~Kim,~\IEEEmembership{Fellow,~IEEE} and Bo~Ai,~\IEEEmembership{Fellow,~IEEE}
\thanks{J. Zheng, J. Zhang and B. Ai are with Beijing Jiaotong University, China; H. Du and D. Niyato are with Nanyang Technological University, Singapore; D. I. Kim is with Sungkyunkwan University, South Korea.}
}

\maketitle
\vspace{-2cm}
\begin{abstract}
The cell-free (CF) massive multiple-input multiple-output (MIMO) system is considered a cutting-edge technology in next-generation mobile communication due to its ability to provide high-performance coverage seamlessly and uniformly.
This paper aims to mitigate the negative impact of channel aging due to the movement of user equipment in CF massive MIMO systems by utilizing rate-splitting (RS) technology.
Taking into account the outdated channel state information, we obtain the achievable sum spectral efficiency (SE) of the RS-assisted CF massive MIMO system, where the private messages can directly adopt conventional maximum ratio, local minimum mean square error (MMSE), and centralized MMSE precoding schemes.
Moreover, we propose a bisection-based precoding scheme that maximizes the minimum SE of common messages, which outperforms the superposition-based and random precoding schemes and exhibits strong robustness in complex mobile scenarios.
Furthermore, we derive a novel closed-form sum SE expression for the considered system.
The results demonstrate that RS technology can mitigate interference in mobile CF massive MIMO systems, improving overall system performance.
\end{abstract}


\IEEEpeerreviewmaketitle

\vspace{-2mm}
\section{Introduction}

The cell-centric architecture in wireless mobile communication networks has inherent limitations in providing uniform and seamless coverage, which impedes the realization of high-performance human-centric service for sixth generation (6G) mobile communications \cite{9390169}. To address this issue, cell-free (CF) massive multiple-input multiple-output (MIMO) has gained significant attention as a promising solution that offers a user-centric networking approach \cite{zhang2020prospective}. The CF massive MIMO system adopts spatial multiplexing to simultaneously serve each user equipment (UE) on the same time-frequency resource with many geographically distributed access points (APs) linked to a central processing unit (CPU) \cite{Ngo2017Cell}.
Therefore, UEs can enjoy reliable and high-rate wireless services regardless of their location within the CF massive MIMO network \cite{8972478,bjornson2019making}.
Importantly, the benefits of CF massive MIMO are highly dependent on efficient transceivers, which in turn require accurate channel state information (CSI) to be obtained in a timely manner \cite{9650567}.

In practice, mobile communication systems may suffer from a phenomenon called channel aging, which is due to UE mobility and results in outdated CSI \cite{zheng2023mobile,truong2013effects}.
This issue is especially challenging in CF massive MIMO systems, as each UE is served by multiple APs, leading to diverse angles of arrival \cite{9322468,9471851}.
Therefore, there has been significant research on investigating the effects of channel aging in CF massive MIMO systems.
In \cite{gao2023hybrid}, a data-knowledge hybrid driven channel semantic acquisition and beamforming solution is proposed for CF massive MIMO systems to overcome the channel aging problem.
One example is our previous work \cite{9322468}, in which we evaluated the impact of channel aging on the performance of CF massive MIMO systems in both uplink and downlink scenarios. Besides, results in \cite{9471851} revealed that channel aging amplifies channel estimation errors and introduces additional interference to CF massive MIMO systems.
Especially, in high mobility scenarios and with long frame durations, the impact of channel aging becomes more significant in CF massive MIMO systems \cite{9406061}.

Recently, the rate-splitting (RS) technique has been devised to mitigate multiuser interference arising from imperfect CSI \cite{7470942}, presenting its great potential for dealing with the problem of channel aging in CF massive MIMO systems. The core concept of RS is to split UE messages into common and private messages, which allows for the partial decoding of interference and the treatment of the remaining interference as noise using successive interference cancellation (SIC) receivers, making it an attractive option for managing interference \cite{9831440}. For instance, the authors of \cite{9737523} introduced a novel RS-assisted downlink transmission framework for CF massive MIMO, which significantly improves SE performance over a conventional CF massive MIMO.
The authors of \cite{10032139} proposed an elegant deep learning RS beamforming for reconfigurable intelligent surface-aided Tera-Hertz massive MIMO systems.
Besides, our previous research \cite{10032129} designed an optimal superposition-based precoding scheme for common messages, which exhibits robustness in asynchronous CF massive MIMO systems.
Importantly, we verified that CF massive MIMO has significant advantage of integration with RS because of its uniform coverage for each UE.
Moreover, the authors in \cite{9491092} first introduced RS into mobile cellular MIMO networks in order to preserve stable multi-user connectivity. As far as we know, none has examined the RS for the CF massive MIMO with channel aging.

Based on the foregoing findings, we look into the achievable sum SE of RS-assisted CF massive MIMO systems with channel aging.
The main contributions are as follows:
\begin{itemize}
  \item Taking into account the outdated CSI, we compare various precoding schemes for private messages, including maximum-ratio (MR), local minimum mean square error (MMSE), and centralized MMSE.
  \item We propose a novel bisection-based precoding scheme for common messages, which significantly outperforms superposition-based and random precoding schemes especially in complex mobile environments.
  \item We derive a closed-form sum SE expression to investigate the effect of inter-user interference and massive antennas on the system performance.
\end{itemize}

\vspace{-2mm}
\section{System model}\label{se:model}

We examine a CF massive MIMO system, where $L$ APs serve all $K$ UEs with different velocities on the same time-frequency resource simultaneously. Each AP is assumed to have $N$ antennas, while each UE is assumed to have a single antenna. Besides, all APs are linked through fronthaul connections to a CPU for joint coherent processing. We consider a resource block consisting of $\tau + 1$ time instants, where the CSI of the time instant $0$ is perfectly known and time instants from $1$ to $\tau$ are used for downlink data transmission.
Then, the channel between AP $l$ and UE $k$ at time instant $n$ (${{\mathbf{h}}_{kl,n}} \in {\mathbb{C}^{N \times 1}}$) can be modelled by Rayleigh fading as \cite{bjornson2019making}
\begin{align}
{{\mathbf{h}}_{kl,n}} \sim \mathcal{C}\mathcal{N}\left( {{\mathbf{0}},{{\mathbf{R}}_{kl}}} \right),n = 0,1, \ldots ,\tau ,
\end{align}
where ${{\mathbf{R}}_{kl}} \in {\mathbb{C}^{N \times N}}$ represents the spatial correlation matrix and ${\beta _{kl}} \triangleq {\text{tr}}\left( {{{\mathbf{R}}_{kl}}} \right)/N$ denotes the large-scale fading coefficient. It is worth noting that due to the impact of UE mobility, ${{\mathbf{h}}_{kl,0}},{{\mathbf{h}}_{kl,1}}, \ldots ,{{\mathbf{h}}_{kl,\tau }}$ are not identical but correlated.

\subsection{Channel Aging}

The phenomenon of channel aging can significantly impact the performance of the system as the estimated CSI gradually becomes outdated over time due to the temporal variations in the propagation environment caused by the relative movement between the APs and UEs. In our considered resource block, the channel for downlink data transmission can be represented as a function of the initial state ${{\mathbf{h}}_{kl,0}}$\footnote{In this paper, we focus on the problem of outdated CSI in the downlink data transmission ($n = 1, \ldots ,\tau$). Therefore, we ignore the effect of channel estimation error and assume the initial state ${{\mathbf{h}}_{kl,0 }}$ is perfectly known. Note that, this assumption does not affect the validity of our findings, and the imperfect CSI caused by channel aging in the channel estimation frame can refer to our previous work \cite{9322468}.} and an innovation component \cite{9322468}, as follows:
\begin{align}
{{\mathbf{h}}_{kl,n}} = {\rho _{k,n}}{{\mathbf{h}}_{kl,0}} + {{\bar \rho }_{k,n}}{{\mathbf{g}}_{kl,n}} , n = 1, \ldots ,\tau ,
\end{align}
where ${{\mathbf{g}}_{kl,n}} \sim \mathcal{C}\mathcal{N}\left( {{\mathbf{0}},{{\mathbf{R}}_{kl}}} \right)$ denotes the independent innovation component at the time instant $n$. Besides, $0 \leqslant {\rho _{k,n}} \leqslant 1$ denotes the time correlation coefficient between time instants $0$ and $n$ for the channel of UE $k$, and ${{\bar \rho }_{k,n}}$ is the error coefficient caused by channel aging, which is calculated as $\sqrt {1 - \rho _{k,n}^2} $. Assuming that the channel varies according to the Jakes' model \cite{chopra2018performance}, we then have
\begin{align}
{\rho _{k,n}} = {J_0}\left( {2\pi {f_{D,k}}{T_s}n} \right),\forall k,\forall n ,
\end{align}
where ${J_0}\left(  \cdot  \right)$ represents the zeroth-order Bessel function of the first kind. The parameter $T_s$ denotes the sampling time, and ${f_{D,k}} = \left( {{v_k}{f_c}} \right)/c$ denotes the Doppler shift experienced by UE $k$. Moreover, ${v_k}$ represents the velocity of UE $k$, $f_c$ represents the carrier frequency, and $c$ represents the speed of light.

\subsection{Outdated CSI}

Despite the assumption of perfect knowledge of CSI in the initial state, the data transmission channel can only get an outdated CSI owing to the time-varying nature of resource blocks. Then, the obtained channel during downlink transmission is expressed as
\begin{align}\label{eq:hhat}
{{{\mathbf{\hat h}}}_{kl,n}} = {\rho _{k,n}{\mathbf{h}}_{kl,0}},n = 1, \ldots ,\tau ,
\end{align}
whose variance is derived as ${\rho ^2_{k,n}}{{\mathbf{R}}_{kl}}$. We can also derive the channel error as
\begin{align}\label{eq:herror}
  {{{\mathbf{\tilde h}}}_{kl,n}} &= {{\mathbf{h}}_{kl,n}} - {{{\mathbf{\hat h}}}_{kl,n}} = {{\bar \rho }_{k,n}}{{\mathbf{g}}_{kl,n}} .
\end{align}
It is clear that the channel error \eqref{eq:herror} is independent of the obtained channel \eqref{eq:hhat}.
Besides, the variance of the channel error can be calculate as
\begin{align}
  {{\mathbf{C}}_{kl,n}} &= \mathbb{E}\left\{ {{{{\mathbf{\tilde h}}}_{kl,n}}{\mathbf{\tilde h}}_{kl,n}^{\text{H}}} \right\} = \mathbb{E}\left\{ {\left( {{{\bar \rho }_{k,n}}{{\mathbf{g}}_{kl,n}}} \right){{\left( { {{\bar \rho }_{k,n}}{{\mathbf{g}}_{kl,n}}} \right)}^{\text{H}}}} \right\} \notag \\
   &= \left( {1 - {\rho ^2_{k,n}}} \right){{\mathbf{R}}_{kl}} .
\end{align}

\newcounter{mytempeqncnt}
\begin{figure*}[t!]
\normalsize
\setcounter{mytempeqncnt}{1}
\setcounter{equation}{9}
\begin{align}
\label{eq:SINRc} {\text{SINR}}_{k,n}^{\text{c}} &= \frac{{{p_{\text{d}}}t{{\left| {{\mathbf{\hat h}}_{k,n}^{\text{H}}{\mathbf{v}}_{{\text{c}},n}^{{\text{norm}}}} \right|}^2}}}{{\frac{{{p_{\text{d}}}\left( {1 - t} \right)}}{K}\sum\limits_{i = 1}^K {{{\left| {{\mathbf{\hat h}}_{k,n}^{\text{H}}{\mathbf{v}}_{i,n}^{{\text{norm}}}} \right|}^2}}  + {p_{\text{d}}}t{{\left( {{\mathbf{v}}_{{\text{c}},n}^{{\text{norm}}}} \right)}^{\text{H}}}{{\mathbf{C}}_{k,n}}{\mathbf{v}}_{{\text{c}},n}^{{\text{norm}}} + \frac{{{p_{\text{d}}}\left( {1 - t} \right)}}{K}\sum\limits_{i = 1}^K {{{\left( {{\mathbf{v}}_{i,n}^{{\text{norm}}}} \right)}^{\text{H}}}{{\mathbf{C}}_{k,n}}{\mathbf{v}}_{i,n}^{{\text{norm}}}}  + {\sigma ^2}}} , \\
\label{eq:SINRp} {\text{SINR}}_{k,n}^{\text{p}} &= \frac{{\frac{{{p_{\text{d}}}\left( {1 - t} \right)}}{K}{{\left| {{\mathbf{\hat h}}_{k,n}^{\text{H}}{\mathbf{v}}_{k,n}^{{\text{norm}}}} \right|}^2}}}{{\frac{{{p_{\text{d}}}\left( {1 - t} \right)}}{K}\sum\limits_{i \ne k}^K {{{\left| {{\mathbf{\hat h}}_{k,n}^{\text{H}}{\mathbf{v}}_{i,n}^{{\text{norm}}}} \right|}^2}}  + \frac{{{p_{\text{d}}}\left( {1 - t} \right)}}{K}\sum\limits_{i = 1}^K {{{\left( {{\mathbf{v}}_{i,n}^{{\text{norm}}}} \right)}^{\text{H}}}{{\mathbf{C}}_{k,n}}{\mathbf{v}}_{i,n}^{{\text{norm}}}}  + {\sigma ^2}}} .
\end{align}
\setcounter{equation}{6}
\hrulefill
\vspace{-0.2cm}
\end{figure*}

\subsection{Rate-Splitting Strategy}

To mitigate the potential impact of outdated CSI on system performance, we employ an RS strategy during downlink data transmission in CF massive MIMO systems considering power constraints at each AP.
RS is achieved by dividing a message of each UE into a common message and a private message, then the common messages are merged into a super common message, and finally all the messages are transmitted simultaneously. At the receiver, the common message is first decoded by each UE with all the private messages being treated as noise. Subsequently, SIC is employed to eliminate the decoded common message from the received signal. Finally, each UE decodes its private message by considering other private messages as noise.
Using normalized precoding vectors $ {\left\| {{\mathbf{v}}_{{\text{c}},l,n}^{{\text{norm}}}} \right\|^2} \leqslant 1$ and $ {\left\| {{{\mathbf{v}}_{il,n}^{{\text{norm}}}}} \right\|^2} \leqslant 1$ for common and private messages in RS strategy, the transmitted signal from AP $l$ at time instant $n$ is derived as
\begin{align}\label{eq:x_signal}
{{\mathbf{x}}_{l,n}} = \sqrt {{p_{\text{d}}}t} {\mathbf{v}}_{{\text{c}},l,n}^{{\text{norm}}}{s_{{\text{c}},n}} + \sqrt {\frac{{{p_{\text{d}}}\left( {1 - t} \right)}}{K}} \sum\limits_{i = 1}^K {{\mathbf{v}}_{il,n}^{{\text{norm}}}{s_{i,n}}} ,
\end{align}
where ${s_{{\text{c}},n}} \!\sim\! \mathcal{C}\mathcal{N}\left( {0,1} \right)$ and ${s_{i,n}} \!\sim\! \mathcal{C}\mathcal{N}\left( {0,1} \right)$ are the common and private messages at time instant $n$, respectively. Besides, ${p_{\text{d}}}$ is the maximum downlink transmit power of each AP. Moreover, the power-splitting factor $0 \! \leqslant \! t \! \leqslant \! 1$ is defined as the proportion of power devoted to the transmission of common messages at each AP. For example when $t \!=\! 0$, the RS-assisted CF massive MIMO system degenerates into a conventional CF massive MIMO system. Note that the power-splitting factors are equal across all APs, and their optimal value to maximize SE can be obtained through our previous research.

\section{Downlink Performance Analysis}\label{se:performance}

In this section, we aim to enhance the achievable sum SE of the RS-assisted CF massive MIMO system under channel aging by designing efficient precoding schemes for common and private messages.

\subsection{Achievable sum SE}

Based on the transmitted signal \eqref{eq:x_signal}, including both common and private messages, the received signal by UE $k$ at time instant $n$ is expressed as
\begin{align}\label{eq:received}
{r_{k,n}} &= \sum\limits_{l = 1}^L {{\mathbf{h}}_{kl,n}^{\text{H}}{{\mathbf{x}}_{l,n}}}  + {w_{k,n}} = \sqrt {{p_{\text{d}}}t} \sum\limits_{l = 1}^L {{\mathbf{h}}_{kl,n}^{\text{H}}{\mathbf{v}}_{{\text{c}},l,n}^{{\text{norm}}}{s_{{\text{c,}}n}}}  \notag\\
 &+ \sqrt {\frac{{{p_{\text{d}}}\left( {1 - t} \right)}}{K}} \sum\limits_{l = 1}^L {{\mathbf{h}}_{kl,n}^{\text{H}}{\mathbf{v}}_{kl,n}^{{\text{norm}}}{s_{k,n}}}  \notag\\
 &+ \sqrt {\frac{{{p_{\text{d}}}\left( {1 - t} \right)}}{K}} \sum\limits_{i \ne k}^K {\sum\limits_{l = 1}^L {{\mathbf{h}}_{kl,n}^{\text{H}}{\mathbf{v}}_{il,n}^{{\text{norm}}}{s_{i,n}}} }  + {w_{k,n}} ,
\end{align}
\begin{thm}\label{coro}
With the help of \eqref{eq:received} and considering the outdated CSI \eqref{eq:hhat}, we can derive the achievable sum SE as\footnote{It is assumed that the UE has knowledge of the obtained channel ${{{\mathbf{\hat h}}}_{kl,n}}$, while the channel error ${{{\mathbf{\tilde h}}}_{kl,n}}$ remains unknown. It is important to note that the obtained channel knowledge can be immediately utilized for signal detection. However, the channel error, with only its distribution known, is considered less useful and is treated as additional interference during signal detection.}
\begin{align}\label{eq:sse}
  {\mathrm{Sum\;}}{{\mathrm{SE}}_n} &= \mathbb{E}\left\{ {{{\min }_k}\left\{ {{\mathrm{SE}}_{k,n}^{\mathrm{c}}} \right\}} \right\} + \sum\limits_{k = 1}^K {\mathbb{E}\left\{ {{\mathrm{SE}}_{k,n}^{\mathrm{p}}} \right\}}  \notag \\
   &= \mathbb{E}\left\{ {{{\log }_2}\left( {1 + {{\min }_k}\left\{ {{\mathrm{SINR}}_{k,n}^{\mathrm{c}}} \right\}} \right)} \right\} \notag \\
   &+ \sum\limits_{k = 1}^K {\mathbb{E}\left\{ {{{\log }_2}\left( {1 + {\mathrm{SINR}}_{k,n}^{\mathrm{p}}} \right)} \right\}} ,
\end{align}
where ${{\mathrm{SINR}}_{k,n}^{\mathrm{c}}}$ and ${{\mathrm{SINR}}_{k,n}^{\mathrm{p}}}$ are respectively given by \eqref{eq:SINRc} and \eqref{eq:SINRp} at the top of this page, with
\setcounter{equation}{11}
\begin{align}
  {{{\mathbf{\hat h}}}_{k,n}} &= {\left[ {{\mathbf{\hat h}}_{k1,n}^{\mathrm{T}}, \ldots ,{\mathbf{\hat h}}_{kL,n}^{\mathrm{T}}} \right]^{\mathrm{T}}} ,  \\
  {\mathbf{v}}_{{\mathrm{c}},n}^{{\mathrm{norm}}} &= {\left[ {{{\left( {{\mathbf{v}}_{{\mathrm{c}},1,n}^{{\mathrm{norm}}}} \right)}^{\mathrm{T}}}, \ldots ,{{\left( {{\mathbf{v}}_{{\mathrm{c}},L,n}^{{\mathrm{norm}}}} \right)}^{\mathrm{T}}}} \right]^{\mathrm{T}}} ,  \\
  {\mathbf{v}}_{i,n}^{{\mathrm{norm}}} &= {\left[ {{{\left( {{\mathbf{v}}_{i1,n}^{{\mathrm{norm}}}} \right)}^{\mathrm{T}}}, \ldots ,{{\left( {{\mathbf{v}}_{iL,n}^{{\mathrm{norm}}}} \right)}^{\mathrm{T}}}} \right]^{\mathrm{T}}} ,  \\
  {{\mathbf{C}}_{k,n}} &= {\mathrm{diag}}\left( {{{\mathbf{C}}_{k1,n}}, \ldots ,{{\mathbf{C}}_{kL,n}}} \right) .
\end{align}
\end{thm}
\begin{IEEEproof}
This result follows similar steps to those used to prove \cite[Theorem 4.1]{bjornson2017massive}, and therefore we omitted it.
\end{IEEEproof}

\subsection{Precoding for Common and Private Messages}

We start by designing precoding vectors for the common messages, with the objective of maximizing the minimum downlink SE of all UEs, subject to power constraints at each AP. It is preferred that all UEs reach the same SE at the optimal solution. Thus, we formulate the following max-min optimization problem:
\begin{align}\label{maximin1}
  \mathop {\max }\limits_{\left\{ {{\mathbf{v}}_{{\text{c}},n}^{{\text{norm}}}} \right\}} \;\mathop {\min }\limits_{k = 1, \ldots ,K} \;&{\text{SE}}_{k,n}^{\text{c}} \notag \\
  \text{s.t.}\;\;\;\;\;&{\left\| {{\mathbf{v}}_{{\text{c}},l,n}^{{\text{norm}}}} \right\|^2} \leqslant 1,\forall l  ,
\end{align}
where ${\text{SE}}_{k,n}^{\text{c}}$ is given in Theorem~\ref{coro}.
Obviously, the problem in \eqref{maximin1} is equivalent to
\begin{align}\label{maxmin2}
  \mathop {\max }\limits_{\left\{ {{\mathbf{v}}_{{\text{c}},n}^{{\text{norm}}}} \right\}} \;\mathop {\min }\limits_{k = 1, \ldots ,K} \;&\frac{{{p_{\text{d}}}t{{\left| {{\mathbf{\hat h}}_{k,n}^{\text{H}}{\mathbf{v}}_{{\text{c}},n}^{{\text{norm}}}} \right|}^2}}}{{{p_{\text{d}}}t{{\left( {{\mathbf{v}}_{{\text{c}},n}^{{\text{norm}}}} \right)}^{\text{H}}}{{\mathbf{C}}_{k,n}}{\mathbf{v}}_{{\text{c}},n}^{{\text{norm}}} + {\kappa _{k,n}}}} , \notag \\
  \text{s.t.}\;\;\;\;\;&{\left\| {{\mathbf{v}}_{{\text{c}},l,n}^{{\text{norm}}}} \right\|^2} \leqslant 1,\forall l ,
\end{align}
where ${{\kappa _{k,n}}}$ is a function of the precoding vector for private message, which is independent of the optimization variable ${{\mathbf{v}}_{{\text{c}},n}^{{\text{norm}}}}$ and is denoted as
\begin{align}
{\kappa _{k,n}} &= \frac{{{p_{\text{d}}}\left( {1 - t} \right)}}{K}\sum\limits_{i = 1}^K {{{\left| {{\mathbf{\hat h}}_{k,n}^{\text{H}}{\mathbf{v}}_{i,n}^{{\text{norm}}}} \right|}^2}}  \notag\\
&+ \frac{{{p_{\text{d}}}\left( {1 - t} \right)}}{K}\sum\limits_{i = 1}^K {{{\left( {{\mathbf{v}}_{i,n}^{{\text{norm}}}} \right)}^{\text{H}}}{{\mathbf{C}}_{k,n}}{\mathbf{v}}_{i,n}^{{\text{norm}}}}  + {\sigma ^2} .
\end{align}
With the help of \cite[Proposition 1]{Ngo2017Cell}, it can be shown that the problem in \eqref{maxmin2} is quasi-concave. Hence, the optimization problem in \eqref{maxmin2} can be effectively addressed by employing the bisection method, which solves a series of convex feasibility problems in each iteration. The detailed algorithm is presented in Algorithm \ref{bisection}.
The number of complex multiplications in one iteration is $\left( {2K + 1} \right)LN + K\left( {{L^2}{N^2} + 1} \right)$, and convergence is fast as the search space is halved with each iteration \cite{bjornson2017massive}.

\begin{algorithm}[t!]
\caption{Bisection Algorithm to Solve \eqref{maxmin2}}
\label{bisection}
\begin{algorithmic}[1]
\Require
Initialize $\gamma_\text{max} \!=\! 10 \max  \! \left\{\! {{\text{SINR}}_{k,n}^{\text{c}}\!\!\left(\! {{\mathbf{v}}_{{\text{c}},l,n}} \!\!=\!\!\! \sum\limits_{i = 1}^K \!{{{\mathbf{v}}_{il,n}}}  \!\!\right)} \!\!\right\}$ and $\gamma_\text{min} \!=\! 0$ to define a relevant range of values for the objective function in \eqref{maxmin2}. Choose a tolerance $\varepsilon > 0$ to determine the stopping criterion.
\Ensure
The percoding vector for common message ${{\mathbf{v}}_{{\text{c}},n}^{{\text{norm}}}}$;
\While{$\gamma_\text{max}$ $-$ $\gamma_\text{min}$ $>$ $\varepsilon$}
\State Set $\gamma = \left(\gamma_\text{max} + \gamma_\text{min}\right)/2$. Solve the following convex feasibility program:
\begin{align}
\left\{ {\begin{array}{*{20}{c}}
  {\sqrt {{p_{\text{d}}}t} {\mathbf{\hat h}}_{k,n}^{\text{H}}{\mathbf{v}}_{{\text{c}},n}^{{\text{norm}}} \geqslant \sqrt \gamma  \left\| {{{\mathbf{u}}_{k,n}}} \right\|} \\
  {{{\left\| {{\mathbf{v}}_{{\text{c}},l,n}^{{\text{norm}}}} \right\|}^2} \leqslant 1,\forall l}
\end{array}} \right. . \notag
\end{align}
\State Besides, ${{\mathbf{u}}_{k,n}}$ is defined as:
\begin{align}
{{\mathbf{u}}_{k,n}} = {\left[ {{{\left( {\sqrt {{p_{\text{d}}}t} {\mathbf{C}}_{k,n}^{\frac{1}{2}}{\mathbf{v}}_{{\text{c}},n}^{{\text{norm}}}} \right)}^{\text{T}}},\sqrt {{\kappa _{k,n}}} } \right]^{\text{T}}} . \notag
\end{align}
\State If problem is feasible, set $\gamma_\text{min} \triangleq \gamma$, else set $\gamma_\text{max} \triangleq \gamma$.
\EndWhile
\end{algorithmic}
\end{algorithm}

\begin{rem}
For reducing the complexity of the bisection-based precoding vector for common message, we propose the superposition-based precoding vector for common message as
\begin{align}\label{Eq:superposition}
{\mathbf{v}}_{{\mathrm{c}},l,n}^{{\mathrm{norm}}} = \sqrt {{\eta _{l,n}}} {{\mathbf{v}}_{{\text{c}},l,n}} = {{{{\mathbf{v}}_{{\mathrm{c}},l,n}}}}/{{\sqrt {\mathbb{E}\left\{ {{{\left\| {{{\mathbf{v}}_{{\mathrm{c}},l,n}}} \right\|}^2}} \right\}} }} ,
\end{align}
where ${{\mathbf{v}}_{{\mathrm{c}},l,n}} = \sum\nolimits_{i = 1}^K {{{\mathbf{v}}_{il,n}}} $ and $\eta _{l,n}$ is the normalized coefficient for common precoding. Previous studies have found that the superposition-based precoding vector for common message performs well when the channels the channels of each UE tend to be asymptotically orthogonal \cite{10032129}.
Furthermore, random precoding is another low-complexity scheme that is commonly used in RS to obtain tractable expressions. This is achieved by setting ${{\mathbf{v}}_{{\mathrm{c}},l,n}} = {{\mathbf{e}}_1}$, where ${{\mathbf{e}}_1}$ is a null vector with a single entry of $1$ and all other entries of $0$ \cite{9491092}.
\end{rem}
\begin{rem}
We observe that the SE expression given in \eqref{eq:sse} is valid for any private precoding vector.
A possible choice is to use the simple MR precoding as
\begin{align}\label{Eq:MR}
{\mathbf{v}}_{kl,n}^{{\mathrm{norm}}} = \sqrt {{\mu _{kl,n}}} {{\mathbf{v}}_{kl,n}} = {{{{\mathbf{v}}_{kl,n}}}}/{{\sqrt {\mathbb{E}\left\{ {{{\left\| {{{\mathbf{v}}_{kl,n}}} \right\|}^2}} \right\}} }} ,
\end{align}
where ${{\mathbf{v}}_{kl,n}} = {{{\mathbf{\hat h}}}_{kl,n}}$ and $\mu _{kl,n}$ is the normalized coefficient for private precoding. It makes the desired signal as strong as possible and costs marginally to compute, but it ignores the possibility of interference.
Besides, if data signal $s_{k,n}$ is decoded at AP $l$ that has local CSI, the local MMSE precoding vector for private message should be used to minimize the MSE, given by
\begin{align}
  {{\mathbf{v}}_{kl,n}} &= \frac{{{p_{\mathrm{d}}}\left( {1 - t} \right)}}{K}\left( {\sum\limits_{i = 1}^K {\frac{{{p_{\mathrm{d}}}\left( {1 - t} \right)}}{K}} } \right. \notag \\
   &\times {\left. {\left( {{{{\mathbf{\hat h}}}_{il,n}}{\mathbf{\hat h}}_{il,n}^{\mathrm{H}} + {{\mathbf{C}}_{il,n}}} \right) + {\sigma ^2}{{\mathbf{I}}_N}} \right)^{ - 1}}{{{\mathbf{\hat h}}}_{kl,n}} .
\end{align}
\end{rem}

\begin{rem}
Moreover, if data signal $s_{k,n}$ is decoded at CPU that owns the CSI of the entire network, the centralized MMSE precoding vector for private message should be used to minimize the MSE, given by
\begin{align}
  {{\mathbf{v}}_{k,n}} &= \frac{{{p_{\mathrm{d}}}\left( {1 - t} \right)}}{K}\left( {\sum\limits_{i = 1}^K {\frac{{{p_{\mathrm{d}}}\left( {1 - t} \right)}}{K}} } \right. \notag \\
  &\times{\left. { \left( {{{{\mathbf{\hat h}}}_{i,n}}{\mathbf{\hat h}}_{i,n}^{\mathrm{H}} + {{\mathbf{C}}_{i,n}}} \right) + {\sigma ^2}{{\mathbf{I}}_{LN}}} \right)^{ - 1}}{{{\mathbf{\hat h}}}_{k,n}} .
\end{align}
Note that, for the centralized MMSE precoding vector, proportional normalization is required to ensure that the association between APs is not severed. Then, we have
\begin{align}
{\mathbf{v}}_{kl,n}^{{\mathrm{norm}}} = {{\mathbf{v}}_{kl,n}}/\sqrt {\mathop {{\mathrm{max}}}\limits_{l = 1, \ldots ,L} \left\{ {\mathbb{E}\left\{ {{{\left\| {{{\mathbf{v}}_{kl,n}}} \right\|}^2}} \right\}} \right\}} .
\end{align}
\end{rem}

\subsection{Closed-form sum SE}

Based on the received signal \eqref{eq:received}, we derive the lower capacity bound using an alternate closed-form formulation called the use-and-then-forget (UatF) bound \cite{Ngo2017Cell,bjornson2017massive}.
\begin{thm}
Considering the MR precoding \eqref{Eq:MR} for private message and the superposition-based precoding \eqref{Eq:superposition} for common message, \mbox{the lower bound on the achievable sum rate is}
\begin{align}
  {\mathrm{Sum\ }}{{\mathrm{SE}}_n} &= {\log _2}\left( {1 + {{\min }_k}\left\{ {{\mathrm{SINR}}_{k,n}^{\mathrm{c}}} \right\}} \right) \notag\\
  &+ \sum\limits_{k = 1}^K {{{\log }_2}\left( {1 + {\mathrm{SINR}}_{k,n}^{\mathrm{p}}} \right)},
\end{align}
with ${\mathrm{SINR}}_{k,n}^{\mathrm{c}} = {{{\mathrm{DS}}_{k,n}^{\mathrm{c}}}}/\left({{{\mathrm{INT}}_{k,n}^{\mathrm{c}} + {\mathrm{INT}}_{k,n}^{\mathrm{p}} + {\mathrm{DS}}_{k,n}^{\mathrm{p}} + {\sigma ^2}}}\right)$ and ${\mathrm{SINR}}_{k,n}^{\mathrm{p}} = {{{\mathrm{DS}}_{k,n}^{\mathrm{p}}}}/\left({{{\mathrm{INT}}_{k,n}^{\mathrm{p}} + {\sigma ^2}}}\right)$, where
\begin{align}
  {\mathrm{DS}}_{k,n}^{\mathrm{c}} &= {p_{\mathrm{d}}}t\rho _{k,n}^4{\left| {\sum\limits_{l = 1}^L {\sqrt {{\eta _{l,n}}} {\mathrm{tr}}\left( {{{\mathbf{R}}_{kl}}} \right)} } \right|^2}, \notag \\
  {\mathrm{INT}}_{k,n}^{\mathrm{c}} &= {p_{\mathrm{d}}}t\sum\limits_{l = 1}^L {{\eta _{l,n}}\sum\limits_{i = 1}^K {\rho _{i,n}^2{\mathrm{tr}}\left( {{{\mathbf{R}}_{il}}{{\mathbf{R}}_{kl}}} \right)} },  \notag \\
  {\mathrm{DS}}_{k,n}^{\mathrm{p}} &= \frac{{{p_{\mathrm{d}}}\left( {1 - t} \right)\rho _{k,n}^4}}{K}{\left| {\sum\limits_{l = 1}^L {\sqrt {{\mu _{kl,n}}} {\mathrm{tr}}\left( {{{\mathbf{R}}_{kl}}} \right)} } \right|^2}, \notag \\
  {\mathrm{INT}}_{k,n}^{\mathrm{p}} &= \frac{{{p_{\mathrm{d}}}\left( {1 - t} \right)}}{K}\sum\limits_{i = 1}^K {\rho _{i,n}^2\sum\limits_{l = 1}^L {{\mu _{il,n}}} {\mathrm{tr}}\left( {{{\mathbf{R}}_{il}}{{\mathbf{R}}_{kl}}} \right)}. \notag
\end{align}
Furthermore, we derive the normalized coefficients as ${\mu _{kl,n}} = 1/\left( {\rho_{k,n}^2{\mathrm{tr}}\left( {{{\mathbf{R}}_{kl}}} \right)} \right)$ and ${\eta _{l,n}} = 1/\left( {\sum\limits_{i = 1}^K {\rho _{i,n}^2{\mathrm{tr}}\left( {{{\mathbf{R}}_{il}}} \right)} } \right)$.
\end{thm}
\begin{IEEEproof}
See Appendix A.
\end{IEEEproof}

\section{Numerical Results and Discussion}\label{se:numerical}

We consider a simulation setting in which $L$ APs and $K$ UEs are distributed uniformly and independently within a square area of $250$ m $\times$ $250$ m. Besides, the three-slope propagation model described in \cite{Ngo2017Cell} is utilized. We focus on communication with a $f_c = 2$ GHz carrier frequency, and the bandwidth is $B =20$ MHz.
Besides, the downlink transmission power is $p_d =23$ dBm and the noise power is $\sigma^2= -96$ dBm \cite{Ngo2017Cell,bjornson2019making}.
Moreover, the sampling frequency is $15$ kHz, and the sampling time is $T_s = 67\;\mu\text{s}$.

\begin{figure}[t!]
\centering
\includegraphics[scale=0.47]{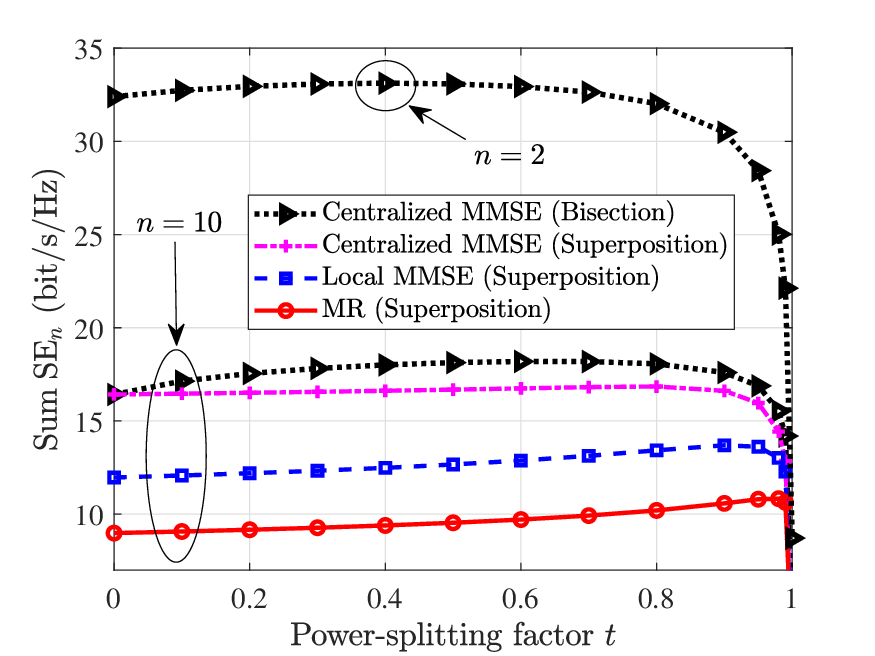}
\caption{Sum SE against the value of power-splitting factor under different precoding schemes ($K=4$, $L=20$, $N=2$, $v_1 = \ldots = v_K = 30$ m/s).
\label{fig:factor}}
\vspace{-0.5cm}
\end{figure}

Fig.~\ref{fig:factor} compares the sum SE against values of the power-splitting factor under different common and private precoding schemes. It is clear that centralized MMSE precoding for private messages is superior to both local MMSE and MR precoding schemes, due to the fact that the global CSI is utilized to mitigate interference significantly. Therefore, the interference elimination brought by RS has little impact on the case of centralized MMSE precoding. Obviously, for the centralized MMSE precoding scheme, the power-splitting factor that achieves the maximum sum SE performance is the lowest compared to the local MMSE and MR precoding schemes. Importantly, our proposed bisection-based precoding for common messages maximizes the potential of RS in enhancing sum SE performance, compared to superposition-based precoding.
Furthermore, an interesting finding from the case using centralized MMSE and bisection-based precoding schemes is that more outdated CSI means more interference, so a larger power-splitting factor is needed to achieve the best sum SE performance.

\begin{figure}[t!]
\centering
\includegraphics[scale=0.47]{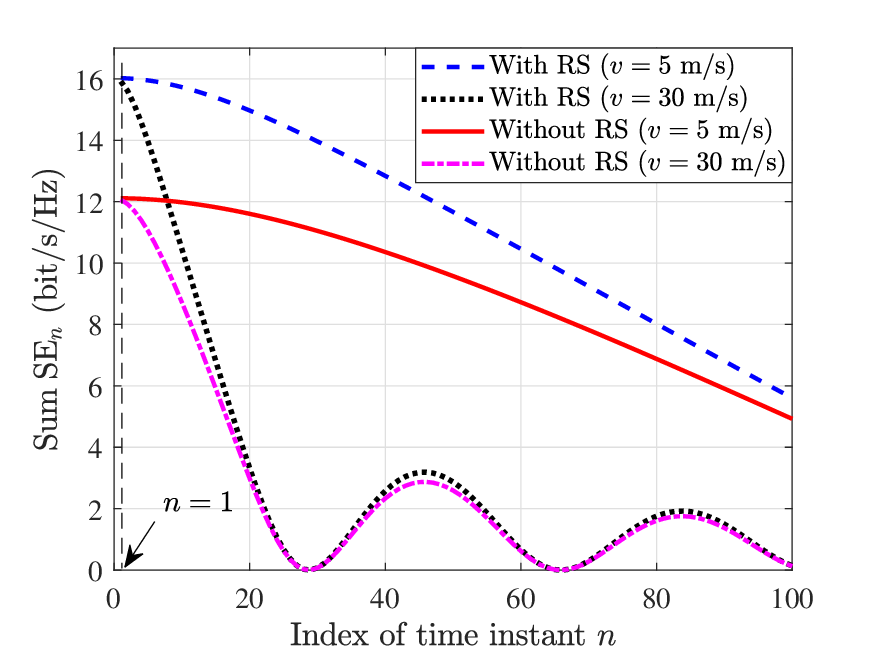}
\caption{Sum SE against the index of time instant ($K=4$, $L=20$, $N=2$, $v = v_1 = \ldots = v_K$).
\label{fig:CAindex}}
\vspace{-0.5cm}
\end{figure}

Fig.~\ref{fig:CAindex} shows the sum SE against the index of time instant with superposition-based precoding for common messages and MR precoding for private messages. It is observed that the sum SE of the system decreases with increasing time instant due to the outdated CSI caused by channel aging, especially when the UE is moving at high velocities. Moreover, the effectiveness of RS is also weakened as CSI becomes outdated.

\begin{figure}[t!]
\centering
\includegraphics[scale=0.47]{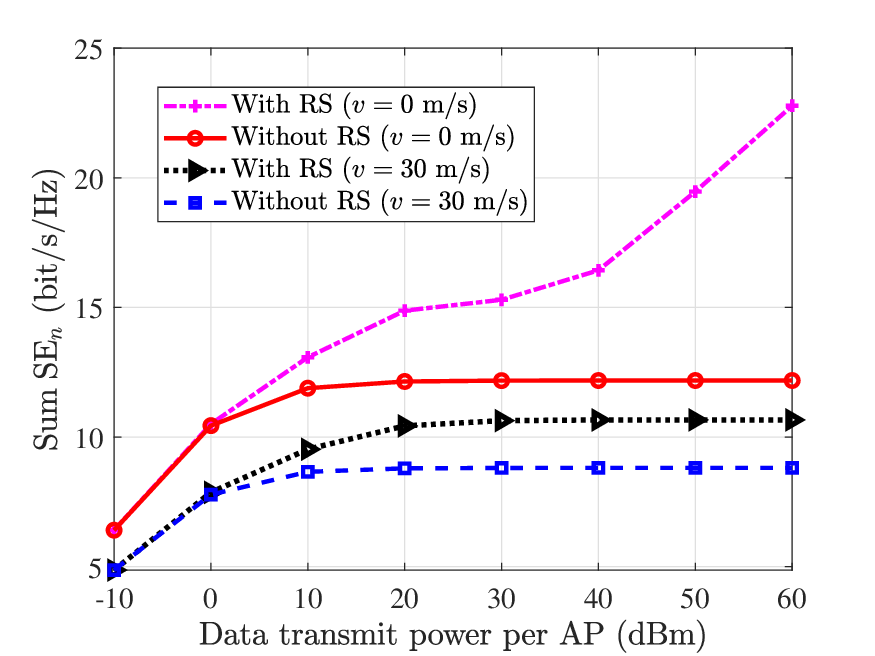}
\caption{Sum SE against the data transmit power per AP ($K=4$, $L=20$, $N=2$, $n=10$, $v = v_1 = \ldots = v_K $).
\label{fig:power}}
\vspace{-0.5cm}
\end{figure}

Fig.~\ref{fig:power} presents the sum SE against the transmit power per AP with superposition-based precoding for common messages and MR precoding for private messages, respectively. As the data transmit power increases, the sum SE without RS reaches a saturation point since the power of the inter-user interference also increases along with the power of the desired signal. Even with the application of RS, the sum SE performance of the system can still saturate due to the residual interference caused by outdated CSI as the UE moves. It is interesting to note that when the UE is stationary, perfect CSI can be acquired, resulting in a continuous improvement in system performance with increasing data transmit power after applying RS.

\begin{figure}[t!]
\centering
\includegraphics[scale=0.47]{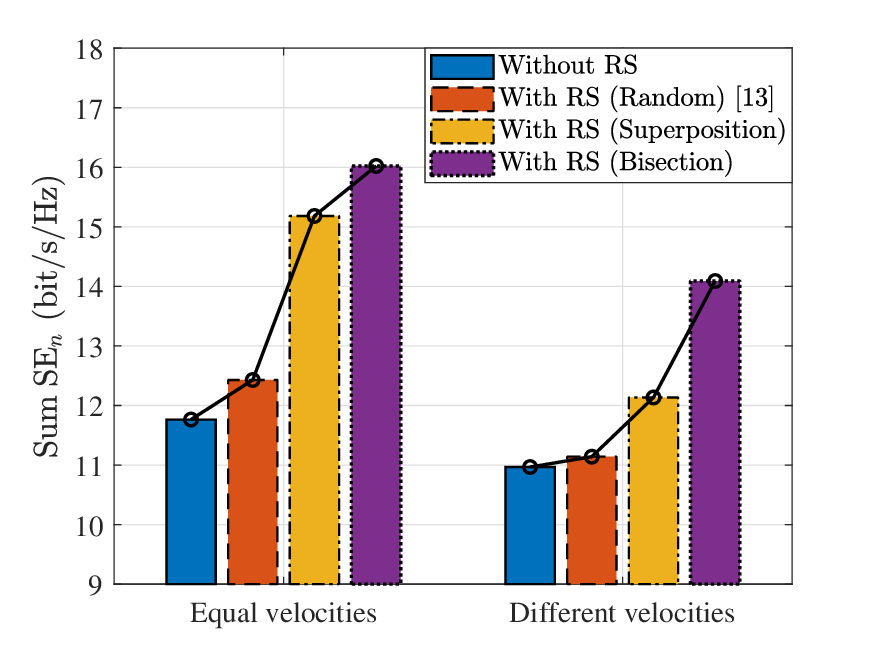}
\caption{Sum SE under different velocity settings ($K=4$, $L=20$, $N=2$, $n=10$).
\label{fig:velocities}}
\vspace{-0.5cm}
\end{figure}

Fig.~\ref{fig:velocities} illustrates the performance improvement of bisection-based, superposition-based and random precoding schemes for common messages under different velocity settings and MR precoding for private messages. We also define two velocity settings, namely equal velocities with $v_i=10$ m/s, $\forall i$ and different velocities with $v_i=0$ m/s, $i \ne k$, $v_k=40$ m/s, $k = \arg \mathop {\min }\limits_i {\left[ {\sum\nolimits_{l = 1}^L {{\beta _{il}}} } \right]_i}$. It is observed that bisection-based precoding achieves the highest SE performance in both equal and different velocity settings. Besides, low-complexity superposition-based precoding approaches the SE performance of bisection-based precoding at equal velocities, but with only a small gain at different velocities.
Moreover, random precoding performs poorly in both cases.
Therefore, the proposed bisection-based precoding scheme is capable of handling various complex mobile scenarios and is considered robust.

\begin{figure}[t!]
\centering
\includegraphics[scale=0.47]{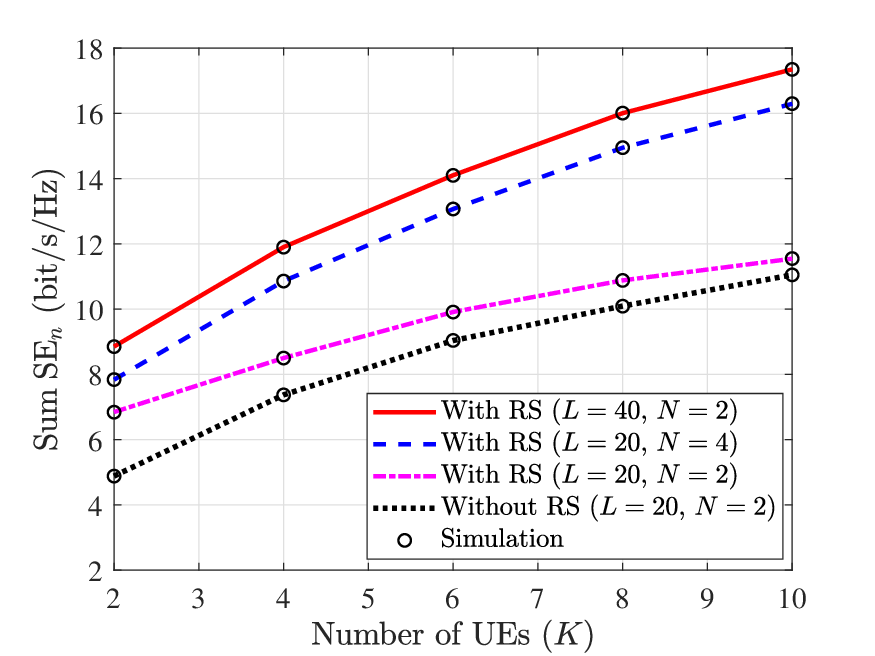}
\caption{Sum SE against different numbers of UEs ($n=10$).
\label{fig:UEs}}
\vspace{-0.5cm}
\end{figure}

Fig.~\ref{fig:UEs} shows that the sum SE increases as the number of UEs increases, but this growth rate gradually slows down due to the inter-user interference. It is also found that the interference suppression capability of RS diminishes as the number of users increases, because the probability of having more poor UEs increases. As expected, increasing the number of APs and antennas per AP significantly enhances system performance thanks to the massive antenna gain achieved.

\vspace{-2mm}
\section{Conclusions}\label{se:conclusions}

This paper focuses on investigating the impact of channel aging resulting from the movement of UE on the downlink transmission of CF massive MIMO systems. To address the issue of interference caused by outdated CSI, we examine the feasibility of employing RS technique. We first derive the achievable sum SE of the RS-assisted CF massive MIMO, where the private messages can adopt arbitrary conventional precoding schemes, such as MR, local MMSE, and centralized MMSE. Importantly, we design a bisection-based precoding scheme to maximize the minimum SE of commom messages, which outperforms the superposition-based and random precoding schemes especially in complex mobile scenarios.
Moreover, we derive a closed-form sum SE expression to study the impact of inter-user interference and massive antennas on the system performance.
In our future work, different power-splitting factors among APs in CF networks will be studied.

\vspace{-0.2cm}
\begin{appendices}
\section{Proof of Theorem 2}
Based on the received signal \eqref{eq:received} and the definition of UatF bound in \cite{bjornson2017massive}, we can derive the values of ${\text{DS}}_{k,n}^{\text{c}}$, ${\text{DS}}_{k,n}^{\text{p}}$, ${\text{INT}}_{k,n}^{\text{c}} $, and ${\text{INT}}_{k,n}^{\text{p}}$ respectively as ${p_{\text{d}}}t{\left| {\sum\limits_{l = 1}^L {\mathbb{E}\left\{ {{\mathbf{h}}_{kl,n}^{\text{H}}{\mathbf{v}}_{{\text{c}},l,n}^{{\text{norm}}}} \right\}} } \right|^2}$,
$\frac{{{p_{\text{d}}}\left( {1 - t} \right)}}{K}{\left| {\sum\limits_{l = 1}^L {\mathbb{E}\!\left\{\! {{\mathbf{h}}_{kl,n}^{\text{H}}{\mathbf{v}}_{kl,n}^{{\text{norm}}}} \!\right\}} } \right|^2}$,
$ {p_{\text{d}}}t\mathbb{E}\!\left\{\! {{{\left| {\sum\limits_{l = 1}^L {{\mathbf{h}}_{kl,n}^{\text{H}}{\mathbf{v}}_{{\text{c}},l,n}^{{\text{norm}}}} } \right|}^2}} \!\right\} \!-\! {\text{DS}}_{k,n}^{\text{c}}$,
and $ \frac{{{p_{\text{d}}}\left( {1 \!-\! t} \right)}}{K}\!\sum\limits_{i = 1}^K \!{\mathbb{E}\!\left\{\! {{{\left| {\sum\limits_{l = 1}^L {{\mathbf{h}}_{kl,n}^{\text{H}}{\mathbf{v}}_{il,n}^{{\text{norm}}}} } \right|}^2}} \!\right\}} \!\! - \!\! {\text{DS}}_{k,n}^{\text{p}}$. The derivation of ${\text{INT}}_{k,n}^{\text{c}} $ is a general case, and thus is provided bellow. With the help of \eqref{Eq:superposition} and \cite[APPENDIX B]{10032129}, we have
\begin{align}
&\mathbb{E}\left\{ {{{\left| {\sum\limits_{l = 1}^L {{\mathbf{h}}_{kl,n}^{\text{H}}{\mathbf{v}}_{{\text{c}},l,n}^{{\text{norm}}}} } \right|}^2}} \right\} = \sum\limits_{l = 1}^L {{\eta _{l,n}}\underbrace {\mathbb{E}\left\{ {{{\left| {\sum\limits_{i = 1}^K {{\mathbf{h}}_{kl,n}^{\text{H}}{{{\mathbf{\hat h}}}_{il,n}}} } \right|}^2}} \right\}}_{{\Upsilon _1}}}  \notag\\
&+\! \sum\limits_{l = 1}^L \!{\sum\limits_{m \ne l}^L \!\!{\sqrt {{\eta _{l,n}}{\eta _{m,n}}} {{\underbrace {\mathbb{E}\!\left\{\! {{\mathbf{h}}_{kl,n}^{\text{H}}\!\sum\limits_{i = 1}^K \!{{{{\mathbf{\hat h}}}_{il,n}}} } \!\right\}}_{{\Upsilon _2}}}^*}\!\!\mathbb{E}\!\left\{\! {{\mathbf{h}}_{km,n}^{\text{H}}\!\sum\limits_{i = 1}^K \!{{{{\mathbf{\hat h}}}_{im,n}}} } \!\right\}} }. \notag
\end{align}
Since channels are independent for different UEs, we derive
\begin{align}
&{\Upsilon _1} = \sum\limits_{i = 1}^K {\mathbb{E}\left\{ {{{\left| {{\mathbf{h}}_{kl,n}^{\text{H}}{{{\mathbf{\hat h}}}_{il,n}}} \right|}^2}} \right\}} = \sum\limits_{i = 1}^K {\mathbb{E}\left\{ {{{\left| {{\mathbf{\hat h}}_{kl,n}^{\text{H}}{{{\mathbf{\hat h}}}_{il,n}}} \right|}^2}} \right\}}  \notag\\
&+\! \sum\limits_{i = 1}^K \!{\mathbb{E}\!\left\{\! {{{\left|\! {{\mathbf{\tilde h}}_{kl,n}^{\text{H}}{{{\mathbf{\hat h}}}_{il,n}}} \!\right|}^2}} \!\right\}}  \!=\!\! \sum\limits_{i = 1}^K \!{\rho _{i,n}^2{\text{tr}}\!\left( {{{\mathbf{R}}_{il}}{{\mathbf{R}}_{kl}}} \right)}  \!+\! \rho _{k,n}^4{\left| {{\text{tr}}\!\left( {{{\mathbf{R}}_{kl}}} \right)} \right|^2} . \notag
\end{align}
Besides, due to ${{{{\mathbf{\hat h}}}_{kl,n}}}$ and $ {{{{\mathbf{\tilde h}}}_{kl,n}}}$ are independent, we have
\begin{align}
{\Upsilon _2} = \mathbb{E}\!\left\{ {{\mathbf{h}}_{kl,n}^{\text{H}}{{{\mathbf{\hat h}}}_{kl,n}}} \right\} \!=\! \mathbb{E}\!\left\{ {{\mathbf{\hat h}}_{kl,n}^{\text{H}}{{{\mathbf{\hat h}}}_{kl,n}}} \right\} \!=\! \rho _{k,n}^2{\text{tr}}\left( {{{\mathbf{R}}_{kl}}} \right) \notag.
\end{align}
Finally, following the similar steps, we can obtain other terms, and this completes the proof.
\end{appendices}

\vspace{0cm}
\bibliographystyle{IEEEtran}
\bibliography{IEEEabrv,Ref}

\end{document}